# An Inkjet Printed Chipless RFID Sensor for Wireless Humidity Monitoring

Michele Borgese, *Student Member, IEEE*, Francesco Alessio Dicandia, *Student Member, IEEE,*
Filippo Costa, *Member, IEEE*, Simone Genovesi, *Member, IEEE,* and Giuliano Manara, *Fellow, IEEE*

*Abstract*— A novel chipless RFID humidity sensor based on a finite Artificial Impedance Surface (AIS) is presented. The unit cell of the AIS is composed of three concentric loops thus obtaining three deep and high-Q nulls in the electromagnetic response of the tag. The wireless sensor is fabricated using low-cost inkjet printing technology on a thin sheet of commercial coated paper. The patterned surface is placed on a metal-backed cardboard layer. The relative humidity information is encoded in the frequency shift of the resonance peaks. Varying the relative humidity level from 50% to 90%, the frequency shift has proven to be up to 270MHz. The position of the resonance peaks has been correlated to the relative humidity level of the environment on the basis of a high number of measurements performed in a climatic chamber, specifically designed for RF measurements of the sensor. A very low error probability of the proposed sensor is demonstrated when the device is used with a 10% RH humidity level discrimination.

*Index Terms*—Radio Frequency Identification (RFID), Chipless RFID, Sensor, Artificial Impedance Surface (AIS).

## I. INTRODUCTION

The research of methods used to transform a chipless RFID tag into a chipless RFID sensor has attracted increasing attention in recent years due to the growing interest in proposing low-cost sensors for realizing the pervasive networks required by the Internet of Things paradigm. At the basis of all proposed solutions there is the use of a material acting as a transducer of the physical parameter that has to be sensed into a change of the RF response of the tag [1], [2]. Different environmental parameters can be monitored depending on the physical properties of the transducer. Although wired sensors are the most popular and well-established ones [1], [3], [4], in recent years wireless sensors based on direct radio-frequency interrogation have been intensively investigated [5], mainly to reduce the price of sensors nodes and deployment cost. One of the first examples is provided in [6], where a chipless RFID temperature sensor is realized by using three functional layers of magnetic materials. An alternative solution is based on Surface Acoustic Wave (SAW) technology. In [7], a wireless temperature and pressure sensor working around 433 MHz is designed by the synergic use of two SAW resonators. In [8], a chipless RFID sensor for detection of $CO_2$ is realized by exploiting the properties of a polymer/single-walled carbon nanotube ink that is used to load one resonator, whilst a second one is used for calibration. Both solutions are fabricated with a Dimatix printer [9] for material deposition. Furthermore, metamaterials have been proposed in [10], [11] for realizing temperature as well as strain sensors. Chipless RFID sensors exploiting the difference between the backscattered signal determined by the structural mode and the antenna mode have been proposed in [12] for monitoring ethylene gas. A sensor based on group-delay and RCS changes has been presented in [13] for humidity monitoring. The Relative Humidity (RH) variation is translated into an alteration of the frequency response by employing silicon nanowires deposited on a PCB chipless RFID tag. In [14], a chipless RFID humidity sensor is realized by adding polyvinyl-alcohol (PVA) on top of a single electric inductive-capacitive resonator. The PVA polymer needs to be dissolved and magnetically stirred before being applied with a droplet on the resonator. In [14], the two antennas are required because the sensing of the humidity level is based on the detection of the transmission coefficient. The shift proves to be up to 607 MHz around 6.87 GHz. As shown in [15], inductive coupling can be used to read chipless RFID humidity sensors. The humidity estimation is based on a frequency shift of 150MHz of the frequency response. The same principle is employed in [16] where a dual-polarized chipless tag is printed with silver nanoparticle-based ink on HP photo paper. This moisture sensor relies on the frequency shift of the peaks encoded in the radar cross section (RCS) of the tag and requires a bandwidth of 13.5 GHz and a dynamic of the RCS within (-80dB,-30dB). Concerning the fabrication methods of printed sensors, standard photolithographic process can be used [17], [18]. An alternative approach is based on the direct printing of the metal pattern with inkjet-printing technology [4], [19]–[23]. The former solution could be problematic since potential air gaps between the resonating particles and the sensitive materials could reduce the sensitivity of measurement. Inkjet-printed sensors are usually fabricated by using a Dimatix material printer, which is suitable for different sensitive materials. The fabrication method is versatile and precise but requires expensive hardware.

The sensor proposed in this paper is instead realized with commercial desktop inkjet printer and the ink does not require any curing. The employed conductive ink is applied to a special paper which acts as a transducer of the humidity level of the external environment into a variation in the frequency response of the chipless RFID tag. The frequency shift has been demonstrated to be as high as 270 MHz without requiring any precise recovering of the RCS value. The sensor is provided with a ground plane (GP), which improves the quality factor of the employed resonator. In addition, the GP allows the sensor to be placed on objects, even metallic ones, without any detuning effect of the frequency response of the tag. Finally, the reading system comprises only a single antenna operating in linear polarization within the bandwidth 2-8 GHz.

The paper is organized as follows. In Section II the periodic configuration of the proposed chipless RFID sensor and the working principle are addressed. Section III aims to properly correlate the electromagnetic response of the tag with the level of humidity of the environment as well as to illustrate the ad-hoc setup designed for the tests. The fabricated prototype and measurements are proposed in Section IV. A statistical analysis of the measured data is also proposed in Section V to assess the reliability and robustness of the chipless sensor. Finally, the conclusions are reported in Section VI.

## II. CHIPLESS RFID SENSOR

The proposed chipless RFID sensor tag is based on a finite Artificial Impedance Surface (AIS). The AIS comprises a finite Frequency Selective Surface (FSS), formed by only a few unit cells, accommodated on a cardboard substrate backed by a metallic ground plane. The FSS is inkjet-printed on a thin sheet of NB-TP-3GU100 Mitsubishi paper. The employed paper is coated with a thin layer composed of polyvinyl alcohol and Aluminum Oxide in order to allow the ink to be correctly deposited on it. The silver nanoparticle ink is deposited with a conventional piezoelectric inkjet printer (Brother DCP-J152W). It is worth highlighting that the conductive pattern is obtained without any heating or sintering of the ink thus obtaining a fast and cost-effective design process [24]. The advantage of this configuration is that the resonators are directly in contact with the substrate and no air gaps are present. Typically, the air gaps invalidate this type of measurements.

A graphic representation of the proposed tag is reported in Fig. 1. The unit cell (P = 15 mm) consists of a three loop FSS printed on a thin paper sheet ($d_m$ = 150 µm) and placed on a thick grounded cardboard layer ($d_c$ = 3 mm) characterized by a permittivity $\varepsilon_r$ = 2.4 - j0.2.

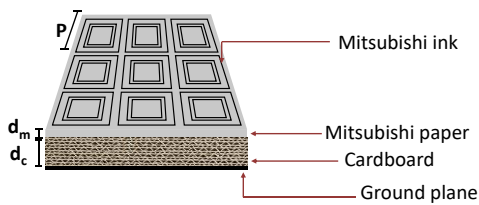

Fig. 1 – Stack-up of the chipless sensor and layout of the resonant element.

### A. Periodic Configuration

The FSS and the ground plane together form a subwavelength resonant cavity. The number of resonances of the cavity is proportional to the resonances of the FSS. If the tag is used on a metallic surface, the ground plane is already available and the tag can be simply realized by printing the FSS on the Mitsubishi paper and gluing it on a cardboard layer, which is then placed on the metal surface. It is worth noting that the reading of a chipless tag composed by a single FSS without the ground plane might be critical. In fact, the presence of objects close to the backside of the tag could affect the reflection response of the FSS, which is a partially transmitting/reflecting surface (PRS). Moreover, the quality factor of the resonances is much higher in the case of metal-backed FSS than for a freestanding FSS. This can be explained either in the frequency domain by using a simple equivalent circuit [25] or in time domain by using the interference theory commonly exploited in the analysis of Fabry-Perot resonators [26]. The latter approach consists of considering the FSS as a partially reflecting mirror. The total reflection coefficient of the cavity is the result of an infinite set of contributions. The first one is due to the first reflection of the FSS, the second one is due to the part of the signal crossing the first FSS that is reflected back by the ground plane, and so on and so forth. A sketch of this time-domain representation is reported in Fig. 2(a). The geometric series converges to the global reflection coefficient of the cavity:

$$R = R_{1a} - \frac{T_1^2 e^{-j2\beta d}}{1 + R_{1b} e^{-j2\beta d}} \quad (1)$$

In (1) $T_1$ represents the transmission coefficient of the FSS, $R_{1a}$ and $R_{1b}$ are the reflection coefficients of the FSS for the two different directions of incidence, which are respectively from the AIS to the GP (AIS→GP) and from the GP to the AIS (GP→AIS). The reflection coefficients $R_{1a}$ is equal to $R_{1b}$ if the FSS is sandwiched between two identical dielectric layers. The propagation constant and the thickness of the cardboard spacer are respectively indicated with β and $d$. As an example, we have simulated the stack-up shown in Fig. 1. In this case, the dielectrics enclosing the FSS are different (air above and photo-paper and thick cardboard below) and consequently $R_{1a}$ and $R_{1b}$ are different. By using those transmission/reflection coefficients, the response of the whole tag including the ground plane can be computed by using relation (1) without repeating the fullwave simulation of the entire structure. In doing that, the phases of the transmission/reflection coefficients have to be preemptively de-embedded to the position of the FSS [27].

The comparison between the interference theory approach and the fullwave simulation of the entire structure preformed with a periodic MoM code (PMM) is reported in Fig. 2(b). The plot shows a good agreement between the two approaches since the FSS is sufficiently distanced from the ground plane and its impedance is not perturbed [28]. If this decoupling condition is not satisfied, the impedance of the FSS is altered by the presence of the ground plane. In this case, the equivalent circuit model would be more accurate than the interference method approach.

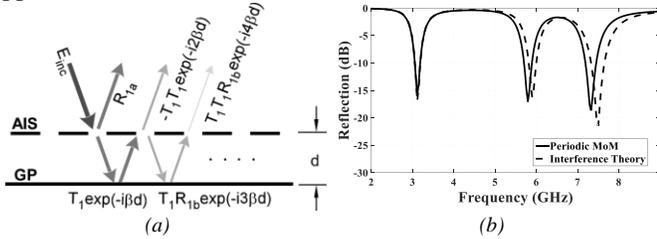

Fig. 2 – (a) Pictorial representation of the reflection coefficient of an AIS; (b) Reflection coefficient of the sensor tag formed by a three loops FSS placed on a 4mm thick cardboard layer backed by a metallic ground plane.

The increasing or decreasing number of unit cells does not improve the coding capacity but it allows increasing or decreasing the level of the RCS and thus improving the tag detection probability for a given reading distance. The effect of the periodic surface on the RCS for a chipless RFID tag based on 5 nested loops printed on top of a 1.6-mm-thick FR4 layer, is shown in Fig. 3. In particular, the RCS as a function of the frequency for a 1x1, 2x2 and 3x3 unit cells is reported. The full wave simulation of the finite structures, performed with a commercial Finite Element Method code (HFSS), shows that by increasing the number of unit cells it is possible to increase both the RCS and the Q-factor of the tag response. Indeed, by comparing the curves reported in Fig. 3, it is possible to observe that the absorption peaks of the 3x3 tag are deeper and sharper with respect to the 1x1 and 2x2 tag. In addition, it is apparent that in the case of the tag composed by a single unit cell, the low frequency peaks between 2GHz and 5GHz are barely recognizable because of the small size of the tag. For completeness, the simulation of the infinite periodic structure performed with the PMM code is also reported in in Fig. 3.

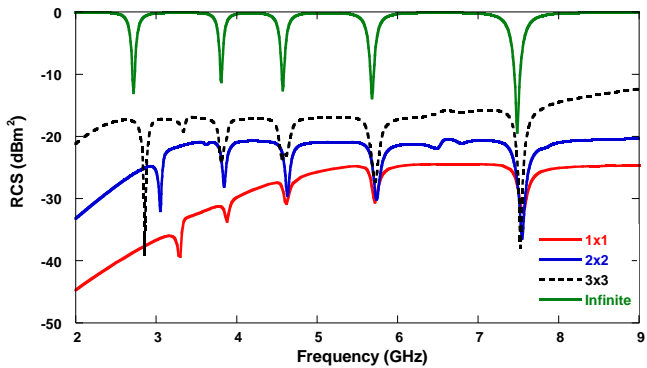

Fig. 3 – Radar Cross Section as a function of the frequency for the chipless RFID tag composed by 1x1 (20 mm x 20 mm), 2x2 (30mm x 30 mm) and 3x3 unit cells (45 mm x45 mm) and for the infinite periodic structure.

B. *Working Principle of the sensor*

The paper substrate with the coating layer acts as a chemical interactive material (CIM) because it is sensitive to humidity variations. For this reason, it has been employed for designing the proposed chipless RFID sensor. The stack-up of the sensor comprises the FSS printed on the Mitsubishi paper which is placed on a metal-backed cardboard substrate. It is worth highlighting that this is the first time that the commercial Mitsubishi ink has been proved to provide humidity sensitivity when deposited on the Mitsubishi paper substrate. This structure exhibits a copolar frequency response with a certain number of deep nulls depending on the topology of the AIS. A variation of the RH level of the surrounding environment produces a variation of the electric permittivity of the CIM with a consequent downshift of the deep nulls of the tag response in the frequency spectrum. The sensing mechanism of the tag is based on the translation of the deep nulls shift into a variation of the RH level in the environment. Even if a single frequency peak is sufficient in principle, the estimation of the RH level at different frequencies with multiple frequency peaks may be beneficial. Indeed, it provides a certain level of redundancy of the relative humidity variations, thus reducing the probability of false reading. Although the cardboard substrate is a moisture sensitive material, the Mitsubishi paper is the only CIM of the sensor because the cardboard layer is not directly exposed to the external environment (and consequently to the moisture). The Mitsubishi paper, the cardboard and the metallic ground plane are bonded together with a plastic adhesive tape, which is placed on the edge of each side of the tag. In addition, the tape isolates the cardboard from the external environment. It is important to highlight the effect of the distance between the resonator elements (the nested rings) and the CIM. If we describe the AIS as a lumped LC circuit, the observed downshift of the resonance frequency is caused by the increase in the aforementioned capacitance value determined by an increased permittivity of the material that is closely interacting with the resonators of the AIS. If the resonators are directly printed on the CIM, there is no air gap and the amount of frequency shift is maximum. On the contrary, it is apparent that even a small air gap can deteriorate the sensor response since it limits the amount of shift achieved with the permittivity change of the CIM. This is clearly reported in Fig. 4 where a double-ring chipless RFID tag with the CIM placed at a distance named '*air gap*' from the resonators is simulated. The stack-up of the analysed structure is shown in the inset of Fig. 4(a). The CIM permittivity varies within [2 - 2.9] and *air gap* within [0 - 0.1] mm. The change of the CIM permittivity leads a remarkable frequency downshift when *air gap* = 0 mm (Fig. 4(a)) but a limited shift is observed when *air gap* = 0.2 mm (Fig. 4(b)).

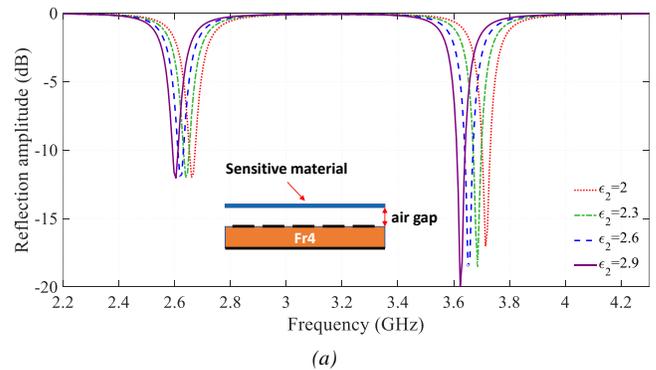

*(a)*

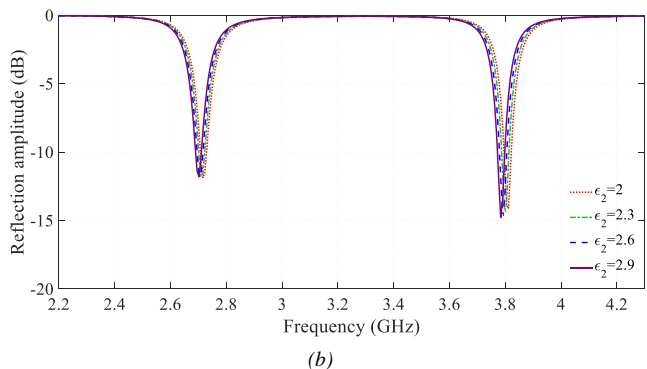

*(b)*

Fig. 4 – Frequency response for a varying CIM permittivity within [2 - 2.9] and *air gap* = 0 *(a) air gap* = 0.2mm *(b)*. Stack-up of the chipless RFID tag with the CIM at distance *air gap* from the resonators in inset *(a)*.

## III. MEASUREMENT SETUP

In order to properly correlate the electromagnetic response of the tag with the level of humidity of the environment, a high number of measurements is required. Consequently, it is important to use a reliable setup that allows the collection of a large amount of data during a certain interval of time. The measurement setup is sketched in Fig. 5.

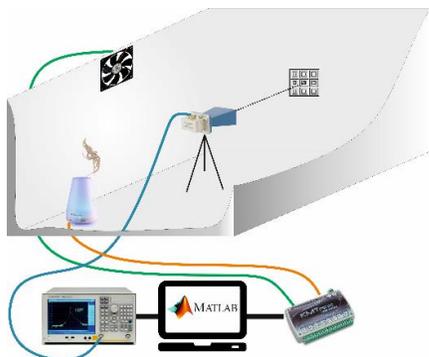

Fig. 5 – Pictorial sketch of measurement setup. The dimension are not in scale.

The chipless tag is attached inside a plastic box of dimensions 77 x 58 x 43cm at a height of 22cm from the bottom of the box. A dual-polarized horn (Flann DP240) is placed in front of the tag at a distance *R* of 25cm and it is connected to a two-ports VNA (Keysight - E5071C). The VNA is connected to a laptop via an USB cable. The electromagnetic response of the tag is measured at the desired time intervals by using a Matlab code. At the same time, it is possible to regulate the level of humidity inside the box with a feedback system controlled by Matlab. To perform this operation, an 8-channel relay (KMTronic USB Relay) driven by the laptop computer is connected to a mist-maker placed inside the box. The relay is also connected to a fan placed inside of the box. The mist-maker can increase the relative humidity (RH) whereas the fan removes the moist air inside the box with a consequent reduction of the RH level. A commercial humidity sensor (RS Pro 1365 Data Logger), connected to the laptop computer, monitors the RH level inside the box. When the RH level exceeds the value defined by the predetermined humidity profile, the fan is activated. In this way, the RH is reduced to the desired level. This Matlab-controlled system composed of the relay, the mist-maker, the fan and the commercial humidity sensor, is able to automatically control the RH level inside the box according to a humidity profile chosen by the user. The humidity profile is defined by a table in which each entry contains the value and the duration of the desired RH level. During the RH controlled cycle, the electromagnetic response of the chipless tag is collected at the predetermined rate. Subsequently, having completed the measurements, the electromagnetic response of the tag is correlated to the RH level with a post-processing algorithm. For the single measure, the correlation is performed on the basis of the timestamp registered by the VNA and the time stamp registered by the Matlab code, which controls the climatic chamber. At the end of this stage, all the data provided by the commercial humidity sensor is associated with the single RF measure.

## IV. PROTOTYPE AND MEASUREMENT RESULTS

The fabricated prototype consists of 3x3 FSS elements placed on a grounded cardboard substrate with thickness equal to 4 mm. The stack-up and the unit cell of the tag are shown respectively in Fig.1 (a) and (b). The periodic pattern has been printed with a water-based silver nanoparticle ink, which has been printed on a single side coated transparency PET film (NB-TP-3GU100 Mitsubishi paper). A stereomicroscope image of the printed chipless tag with a scale bar of 200 µm is shown in Fig. 6a. An enlarged version of this picture is reported in Fig. 6b in which the scale bar is 500 µm. Despite the use of a commercial printer, the Fig. 6(b) shows a very high accuracy of the printing process. The area between one loop and the other is clean and there is no presence of satellite drops.

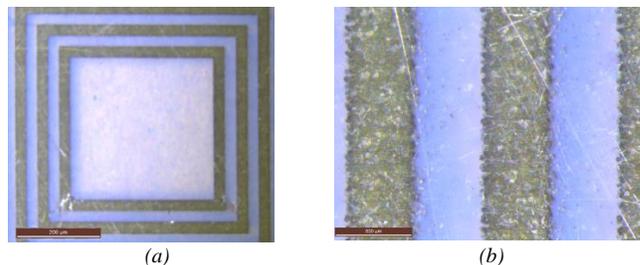

*(a)* *(b)*

Fig. 6 – Stereomicroscope image of the printed chipless tag. The scale bar is 200 µm in (a) and 500 µm in (b). The microscope adopted in this work is a Leica M165C.

After carrying out the measurements and the post processing stage, it is possible to link the variation of the RH level in the climatic chamber to the variation of the electromagnetic response of the proposed chipless RFID tag. The frequency responses of the tag measured for RH levels, equals to 50%, 70%, 80% and 90% are shown in Fig. 7. It is evident that the tag response exhibits 3 deep nulls at the frequencies of 3.076 GHz, 5.888 GHz and 7.275 GHz when RH=50% (minimum RH level achievable in the climatic chamber). By increasing the RH level inside the box, a downshift of the resonance peaks for the three frequencies is visible. This is due to the fact that the paper absorbs the moisture inside the climatic chamber and increases its electrical permittivity. Consequently, the resonant frequencies of the printed tag decrease. The higher the RH level the greater the downshift of the frequency response of the tag. The cardboard does not suffer from any degradation when the RH level is in the range 50% to 90% at room temperature. In addition, as mentioned in Section II, the cardboard layer is not directly exposed to the external environment. The Mitsubishi paper employed for the fabrication of the tag is a PET film which does not degrade with the humidity of the external environment. According to the datasheet [29], the ink exhibits a good solubility in water.

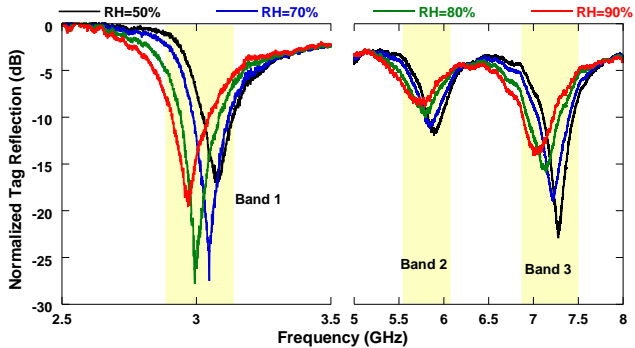

Fig. 7 – Normalized reflection coefficient measurement of the chipless tag for different humidity levels.

In order to verify whether the performance of the sensor is compromised when it is employed in extreme conditions of humidity and if the process is reversible, the tag has been repeatedly treated with nebulized water. The tag response as a function of the frequency for different condition is shown in Fig. 8.

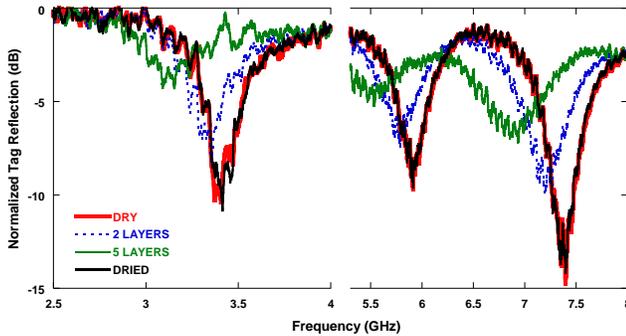

Fig. 8 – Tag response as a function of the frequency for different conditions of the tag. Several layers of nebulized water are deposited on the surface of the tag. Subsequently, the tag is dried with warm air.

After the deposition of two layers of nebulized water, a down shift and a reduction of the depth of the absorption peaks is observed. Subsequently, once two additional layers of nebulized water are deposited, the frequency response of the tag is completely degraded. Indeed, the high frequency peak is still visible and a further downshift of the peak is present. Conversely, the other two peaks are not recognizable. It is worthwhile to point out that this extreme test condition is outside the considered working range and much more severe than 90% humidity. However, it is interesting to observe that, once the tag is dried with warm air, the frequency response of the tag returns to its initial state. As shown in Fig. 8, the curves of the dry and dried tag are superimposed. From this result, it is possible two conclude that the tag is able to absorb and release humidity with a reversible process. In addition, although the ink exhibits a good solubility in water, the direct application layers of water on the surface of the tag, does not permanently jeopardize the function of the sensor.

In order to realize a humidity sensor with the proposed chipless tag, it is necessary to correlate the position of the resonance peaks to the RH level. To perform this task, the frequency response of the tag was monitored when the RH level within the climate chamber was increased from 60% to 90% in steps of 10%. To test the moisture absorption time of the tag, each humidity level was kept constant for 15 minutes. Furthermore, to the purpose of verifying whether the paper was able to release the absorbed moisture, after reaching the 90%, the RH level was decreased to 60% in steps of 10%, similar to the previous process. In this way, a stepped RH profile with a total observation equal to 105 minutes is obtained. The measurements of the tags were carried out at intervals of 10 seconds, thus obtaining 630 measures in the observation time. A short measurement interval has been chosen in order to appreciate the reaction time of the tag to the variation of the RH level.

With the aim of better displaying the shift of the resonance peaks with the variation of the RH level, the resonant frequencies as a function of the observation time has been plotted. The curves show the behavior of all three working frequencies of the tag (Fig. 9 (a), (b), (c)). The level of humidity inside the climatic chamber as a function of the observation time is superimposed in the same graphs. In all three of the plots, it is evident that in correspondence to a rapid variation of the RH curve, the resonance frequencies vary rapidly as well. On the contrary, when the RH level is constant, the resonance frequencies are almost constant. In addition, these graphs confirm that the moisture absorption of the tag is a reversible phenomenon. In fact, when the RH level is decreased, the frequency of the three deep nulls increases, thus obtaining the symmetric RH profile shown in Fig. 9 (a), (b), (c).

In order to compare the magnitude of the frequency shift for the three bands, it is convenient to consider the normalized resonance frequency shift.

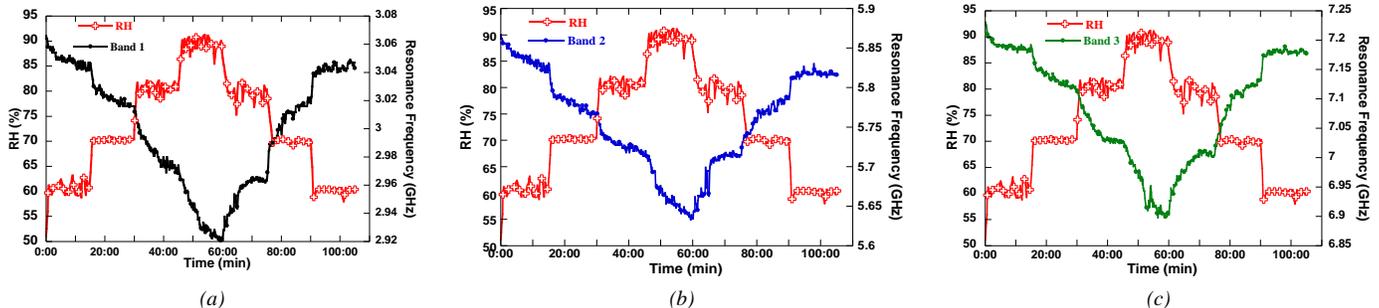

Fig. 9 – RH level and resonance frequency as a function of the observation time for the three working bands of the chipless tag: Band 1 (a), Band 2 (b) and Band 3 (c).

It is defined as the value of the resonance frequency at a certain time, $f_0(t)$, divided by the value of the resonance frequency registered at the time $t_0$ when the minimum level of RH is present in the chamber:

$$NormFreq(t) = \frac{f_0(t)}{f_0(t_0)} \quad (2)$$

In this way, it is possible to calculate the percentage frequency shift with respect to the resonance frequency measured in normal humidity conditions. The RH level and the normalized frequency shift as a function of the observation time for the three working bands of the chipless tag is reported in Fig. 10.

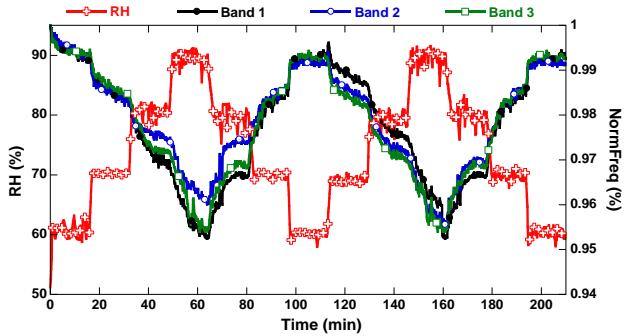

Fig. 10 – RH level and the normalized resonance frequency shift as a function of the observation time for the three working bands of the chipless tag.

It is interesting to note that the tag exhibits almost the same relative shift in the three bands. This property is very useful because it increases the redundancy of the information retrieved by the interrogation of the chipless tag with a consequent reduction of the false reading probability of the RH level.

## V. STATISTICAL ANALYSIS

In order to analyse the reliability and robustness of the proposed sensor, a statistical analysis of the measured data has been performed. The resonant frequencies collected from the electromagnetic measures can be considered as a set of random variables with a certain probability density function (PdF). In particular, the PdF for different RH levels for the three working bands have been calculated and are respectively shown in Fig. 11 (a), (b), (c). The measured frequency response of the tag is affected both by the oscillations of the RH level inside the climatic chamber and by some inaccuracies of the designed sensor. Ideally, in absence of errors introduced by the environment and equipment, the PdF for all resonance frequencies would be a Dirac delta function. However, because of measurement uncertainties, the statistical distribution of the measurement is not a Dirac delta function. It is demonstrated that the noise introduced by the environment, the equipment and the sensor can be modelled as *Additive White Gaussian Noise* (AWGN). To this purpose, in Fig. 11 the Normal PdF with the same mean value and variance calculated from measured data have been reported. The figure shows a good agreement between the measured data and the Normal distributions. In the graphs of the measurements, some fluctuations around the desired value in the stepped RH profile are visible. These fluctuations are due to the feedback mechanism used to stabilize the humidity in the climatic chamber. In addition, these fluctuations produce additional residual uncertainty in the measurements with a consequent increase of the variance of the PdF. The data collected from these graphs has been reported in Table I. In the table, the mean value ($\eta$), the variance ($\sigma^2$) and the maximum distance of the resonance frequency with respect to the mean value ($\Delta_{max}$), have been reported for different RH levels and for the three working bands of the chipless tag. Based on the Gaussian model of the noise, it is interesting to evaluate the performance of the tag when it is employed as a threshold humidity sensor. In particular, the error probability ($P(e)$) in the estimation of RH level by using a single interrogation of the tag can be calculated as follows:

$$P(e) = \sum_{n=1}^{N} P(e \mid RH = RH_{k(n)}) P(RH = RH_{k(n)}), k = 60, 70, 80, 90 \quad (3)$$

where $k$ represents the set of the RH levels (sample space of the variable RH) and $N$ is the number of RH levels assumed by $k$. $P(RH = RH_{k(n)})$ represents the a-priori probability that RH assumes the value $RH_{k(n)}$ and $P(e \mid RH = RH_{k(n)})$ are the conditional probabilities. In this model, the RH level is assimilated to a discrete random variable with uniform distribution with the a-priori probabilities equal to $1/N$. In this case, the optimum thresholds ($\lambda_n$) are fixed in the middle of the mean values (Fig. 12) according to the maximum a-posteriori probability (MAP).

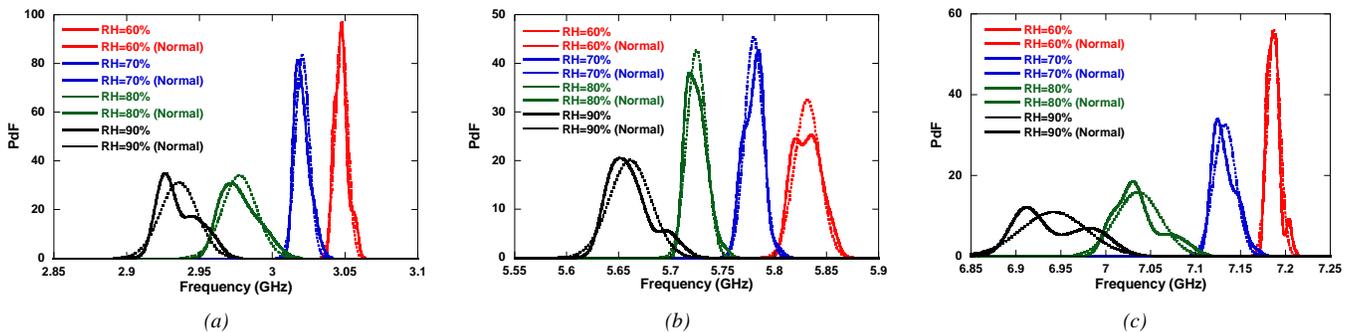

Fig. 11 – Comparison between the PdF of the resonance frequency and the Normal PdF for different RH levels (60%, 70%, 80%, 90%) for the three working bands of the chipless tag: Band 1 (a), Band 2 (b) and Band 3 (c).

Table I - Statistical analysis of the performance of the climate chamber.

|  |  | Band 1 | Band 2 | Band 3 |
|---|---|---|---|---|
| **RH=60%** | η (GHz) | 3.047 | 5.831 | 7.186 |
|  | $\sigma^2$ | 1.83e-5 | 1.50e-4 | 5.02e-5 |
|  | $\Delta_{max}$ (GHz) | 0.010 | 0.024 | 0.023 |
| **RH=70%** | η (GHz) | 3.020 | 5.780 | 7.132 |
|  | $\sigma^2$ | 2.28e-5 | 7.69e-5 | 1.49e-4 |
|  | $\Delta_{max}$ (GHz) | 0.014 | 0.026 | 0.036 |
| **RH=80%** | η (GHz) | 2.977 | 5.724 | 7.036 |
|  | $\sigma^2$ | 1.37e-4 | 8.69e-5 | 6.32e-4 |
|  | $\Delta_{max}$ (GHz) | 0.028 | 0.029 | 0.064 |
| **RH=90%** | η (GHz) | 2.935 | 5.661 | 6.942 |
|  | $\sigma^2$ | 1.63e-4 | 3.95e-4 | 13e-4 |
|  | $\Delta_{max}$ (GHz) | 0.028 | 0.053 | 0.074 |

Consequently, in order to evaluate the $P(e)$, the following conditional probabilities are needed:

$$P(e|RH=60\%) = \int_{-\infty}^{\lambda_3} f(x|RH=60\%)dx$$
$$P(e|RH=70\%) = \int_{-\infty}^{\lambda_2} f(x|RH=70\%)dx + \int_{\lambda_3}^{\infty} f(x|RH=70\%)dx \quad (4)$$
$$P(e|RH=80\%) = \int_{-\infty}^{\lambda_1} f(x|RH=80\%)dx + \int_{\lambda_2}^{\infty} f(x|RH=80\%)dx$$
$$P(e|RH=90\%) = \int_{\lambda_1}^{\infty} f(x|RH=90\%)dx$$

Once the PdF shown in Fig. 11 are known, it is possible to calculate the error probability $P(e)$ in the estimation of RH level with a single interrogation of the tag for each working frequency band (Table II). The overall error probability of the sensor could be further reduced with a thorough processing of the received signal. This approach is widely used in *Multiple Input Multiple Output* (MIMO) applications in which the presence of multiple transmitting and receiving antennas allows the reliability of the system to be improved. In these applications, the spatial diversity provides a redundancy of the received information, which can be exploited by techniques such as the Maximum Ratio Combining (MRC) [30] to reduce the error probability of the MIMO link. It is important to remark that in the case of the presented sensor, the redundancy of the information is provided by the frequency diversity.

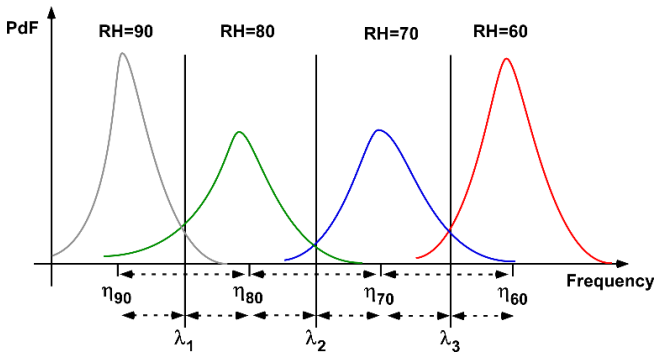

Fig. 12 – Graphical representation of the optimal thresholds for the MAP decision criterion in the case of Gaussian distribution.

Table II – $P(e)$ in the estimation of RH level with a single interrogation of the tag for each working band.

|  | Band 1 | Band 2 | Band 3 |
|---|---|---|---|
| $P(e)$ | 0.031 | 0.019 | 0.042 |

## VI. CONCLUSION

A novel chipless RFID humidity sensor based on an Artificial Impedance Surface chipless tag has been presented. The tag comprises a cardboard substrate and a thin sheet of Mitsubishi paper with the FSS resonators printed on top. The conductive ink has been printed with a commercial piezoelectric printer without any curing or sintering of the ink thus obtaining a very fast and cost effective printing process. In order to correlate the RH level of the external environment, a high number of measurements of the sensor have been performed in a Matlab-driven climatic chamber designed specifically. The estimation of the RH level of the environment is based on the normalized frequency shift of the AIS resonance peaks. The results have shown a high sensitivity of the chipless tag to the small variation of the RH level. Indeed, changing the relative humidity level from 50% to 90%, the frequency shift has proven to be up to 270 MHz. In addition, the statistical analysis of the robustness and reliability of the proposed chipless RFID tag when it is employed as a threshold sensor has been presented. The statistical analysis performed on the collected data showed that the tag exhibits a low error probability when employed as a threshold sensor.

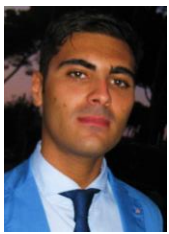

**Michele Borgese** received B.E. and M.E. degrees in telecommunications engineering from the University of Pisa, Italy, in 2010 and 2013, respectively. In 2017, he received the PhD degree in Ingegneria dell'Informazione from the University of Pisa where he is currently a Post-Doctoral Researcher at the Microwave and Radiation Laboratory. Currently, he is involved in the design of multiband reflactarrays and chipless RFID sensors.

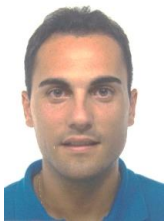

**Francesco Alessio Dicandia** received the Bachelor's and Master's degrees in telecommunications engineering from the University of Pisa, Pisa, Italy, in 2012 and 2014, respectively, where he is currently pursuing the Ph.D. degree. His current research interests include reconfigurable antennas, multiple-input and multiple-output antennas, non-Foster matching network, and characteristic modes analysis.

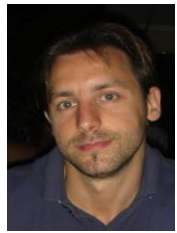

**Filippo Costa** received the M.Sc. degree in telecommunication engineering and the Ph.D. degree in applied electromagnetism from the University of Pisa, Pisa, Italy, in 2006 and 2010, respectively. In 2009, he was a Visiting Researcher at the Department of Radio Science and Engineering, Helsinki University of Technology, TKK (now Aalto University), Finland. During the period 2015-2017, he was several times a short-term Visiting Researcher at Grenoble Institute of Technology, Valance, France and at University Rovira I Virgili, Tarragona, Spain.

He is currently an Assistant Professor at the University of Pisa. His research interests include metamaterials, metasurfaces, antennas and Radio Frequency Identification (RFID).

He serves as an Associate Editor of the IEEE SENSORS LETTERS. In 2015 and 2016, he was appointed as outstanding reviewer of IEEE TRANSACTIONS ON ANTENNAS AND PROPAGATION. He was recipient of the Young Scientist Award of the URSI International Symposium on Electromagnetic Theory, URSI General Assembly and URSI AT-RASC in 2013, 2014 and 2015, respectively.

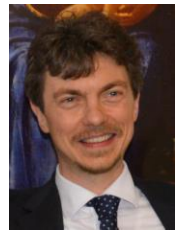

**Simone Genovesi** (S'99-M'07) received the Laurea degree in telecommunication engineering and the Ph.D. degree in information engineering from the University of Pisa, Pisa, Italy, in 2003 and 2007, respectively. Since 2003 he has been collaborating with the Electromagnetic Communication Laboratory, Pennsylvania State University (Penn State), University Park. From 2004 to 2006 he has been a research associate at the ISTI institute of the National Research Council of Italy (ISTI-CNR) in Pisa. From 2010 to 2012 he has been also a postdoctoral research associate at the Institute for Microelectronics and Microsystems of the National Research Council of Italy (IMM-CNR).

He is currently an Assistant Professor at the Microwave and Radiation Laboratory, University of Pisa. He is also a member of RaSS National Laboratory of CNIT (Consorzio Nazionale Interuniversitario per le Telecomunicazioni). Current research topics focus on metamaterials, radio frequency identification (RFID) systems, optimization algorithms and reconfigurable antennas.

He was the recipient of a grant from the Massachusetts Institute of Technology in the framework of the MIT International Science and Technology Initiatives (MISTI).

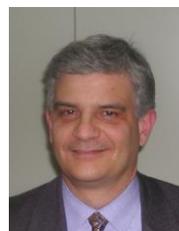

**Giuliano Manara** received the Laurea (Doctor) degree in electronics engineering (summa cum laude) from the University of Florence, Italy, in 1979. Currently, he is a Professor of the University of Pisa, Italy. From 2000 to 2010 and since 2013, he has been serving as the President of the Bachelor and the Master Programs in Telecommunications Engineering at the same University. Since 2010, he has been serving as the President of the Bachelor Program in Telecommunications Engineering at the Italian Navy Accademy in Livorno, Italy. Since 1980, he has been collaborating with the Department of Electrical Engineering of the Ohio State University, Columbus, Ohio, USA, where, in the summer and fall of 1987, he was involved in research at the ElectroScience Laboratory.

His research interests have centered mainly on the asymptotic solution of radiation and scattering problems. He has also been engaged in research on numerical, analytical and hybrid techniques, frequency selective surfaces (FSS) and electromagnetic compatibility. More recently, his research has also been focused on antenna design and on Radio Frequency Identification (RFID).

Prof. Manara was elected an IEEE (Institute of Electrical and Electronic Engineers) Fellow in 2004 for "contributions to the uniform geometrical theory of diffraction and its applications.". He served as the International Chair of URSI Commission B for the triennium 2011-2014. Prof. Manara has served as the General Chair of the International Symposium on Electromagnetic Theory (EMTS 2013), held in Hiroshima, Japan, May 20-24, 2013.

Prof. Manara is the President of CUBIT (Consortium UBIquitous Technologies) S.C.A.R.L., a consortium created by the Dipartimento di Ingegneria dell'Informazione of the University of Pisa, Polo Navacchio S.p.A. (Navacchio, Cascina) and some highly innovative Italian companies, with the aim of defining a new knowledge transfer model from university to industry.